\documentstyle[12pt]{article}
\begin{document}
\title{Deuteron Electromagnetic Form Factors \\
in the Intermediate Energy Region }
\author{
{Jun Cao$^{1)}$ and  Hui-fang Wu$^{2,1)}$} \\
{\small 1) Institute of High Energy Physics, P.O. Box 918(4), Beijing,
100039 People's Repulic of China\thanks{Mailing address.} }\\
{\small 2) CCAST (World Laboratory) P.O. Box 8730, Beijing, 100080,
People's Repulic of China}
}
\date{}
\maketitle
\vskip 0.8in
\begin{abstract}

\begin{quotation}
\noindent
Based on a Perturbative QCD analysis of the deuteron form factor, a model
for the reduced form factor is suggested. The numerical result is
consistent with the data in the intermediate energy region.\\
\vskip 5pt
\noindent
PACS number(s):  13.40.Gp, 12.38.Bx, 24.85.+p, 27.10.+h
\end{quotation}
\end{abstract}
\vskip 0.6in
\section*{I. INTRDUCTION}
\indent
\par
Exclusive processes involving the hadron at large momentum transfer were first
studied in perturbative QCD(pQCD) by Brodsky and Lepage \cite{bl}.
Analysis of the deuteron form factor in the intermediate energy region
 \cite{ks} revealed that QCD could strongly affect the behavior of the 
deuteron electromagnetic form factors when $Q^2$ 
is the order of several GeV$^2$.
It was pointed out in Refs.~\cite{bh,coester} that the domain for leading-power
pQCD predictions for the deuteron form factors is 
$Q^2 \gg 2M_{d}{\Lambda}_{QCD}\sim 0.8{\rm GeV}^2 $ where a calculation 
with the Paris potential \cite{cckp} shows explicit deviation, although it can explain 
the data well for $Q^2 < 1 {\rm GeV}^2$(See Fig.~1).
In this domain the deuteron 
form factor can be written to the leading order in $1/Q^2$ as a convolution: 
\begin{equation}\label{fdeq}
F_d(Q^2)=\int_0^1 [dx][dy] \Phi_d^{\dag}(y_j,Q)
T_H^{6q+\gamma^{\ast}\rightarrow 6q}(x_i,y_j,Q) \Phi_d(x_i,Q)\hskip 0.1in ,
\end{equation}
where the distribution amplitude $\Phi_d(x_i,Q)$ is defined as \cite{dy}
\begin{equation}
\Phi_d(x_i,Q)=\int^{Q} [d^2{\bf k}_{\perp}] \Psi_{6q/d}(x_i,
{\bf k}_{\perp i})\hskip 0.1in .
\end{equation}
The notation $[dx]$ and $[d^2 {\bf k}_{\perp} ]$ is denoted by
\begin{eqnarray}
[dx]  &\equiv& \delta (1- \sum_{i=1}^n x_i) \prod_{i=1}^{n} dx_i 
\hskip 0.1in ,\\\mbox{}
[d^2 {\bf k}_{\perp}] &\equiv& 16{\pi}^3 \delta(\sum\limits_{i=1}^n 
      {\bf k}_{\perp i} )
\prod\limits_{i=1}^n \frac{d^2 {\bf k}_{\perp i}}{16{\pi}^3} \hskip 0.1in .
\end{eqnarray}
However, the calculation of the normalized
$T_H^{6q+\gamma^{\ast}\rightarrow 6q}$
to leading order in $\alpha_{s}(Q^2)$ would require the evaluation of over
300,000 Feynman diagrams involving five gluons. Farrar {\it et al.}
 \cite{fhz} have done perturbative calculations on the helicity zero to zero
deuteron form factor and found that it is much smaller than the deuteron 
form factor data at experimentally accessible momentum transfer.
\par
In order to make more detailed and experimentally accessible predictions,
it was suggested in
Ref.~\cite{bh} to define a reduced nuclear form factor by removing
the nucleon compositeness,
\begin{equation}\label{redu}
f_d(Q^2) \equiv \frac{F_d(Q^2)} {F_N^2(Q^2/4)}\hskip 0.1in .
\end{equation}
The argument for each of the nucleon form factors, $F_N$, is $Q^2/4$ since,
in the limit of zero binding energy, each nucleon must change its momentum
from $P/2$ to $(P+q)/2$.
\par
For the reduced form factor of the deuteron one finds the asymptotic
scaling behavior \cite{bjl}
\begin{equation}\label{scal}
Q^2 f_d(Q^2)\sim \left(ln \frac{Q^2}{\Lambda^2}
               \right)^{-1-\frac{2C_F}{5\beta}} \hskip 0.1in ,
\end{equation}
which can be compared with the available data in the large $Q^2$ region,
although this prediction is only for asymptotic momentum transfer.
Equation~(\ref{scal}) reminds us the reduced
form factor of deuteron may be derived in a way which 
is similar to the meson case in a perturbative
QCD calculation. 
\par
The aim of this paper is to build a model to calculate the reduced form
factor of the deuteron in the intermediate energy region.
The remainder of this paper is organized as follows:
In Sec.~II we analyze the reduced form factor, $f_d(Q^2)$, and the
deuteron wave function. A QCD inspired model
is built in Sec.~III. In Sec.~IV
the numerical results for $f_d(Q^2)$ is given from our model. The
final section is reserved for summary and discussion.

\section*{II. THE REDUCED FORM FACTOR AND THE WAVE FUNCTION OF THE DEUTERON}
\indent
\par
In the case of electron-deuteron elastic scattering, the standard Rosenbluth
cross section \cite{hof} is written (in the laboratory frame) as
\begin{equation}
\frac{d\sigma}{d\Omega}=(\frac{d\sigma}{d\Omega})_{Mott}
[A(Q^2)+B(Q^2)\tan^2(\frac{\theta}{2})]\hskip 0.1in ,
\end{equation}
where $A(Q^2)$ and $B(Q^2)$ are determined by $G_C$, $G_M$ and 
$G_Q$ \cite{acg}:
\begin{eqnarray}
A(Q^2) & = & G_C^2+\frac{2}{3} \eta G_M^2+\frac{8}{9}\eta^2 G_Q^2
\hskip 0.1in  ,\\
B(Q^2) & = & \frac{4}{3}\eta (1+\eta )G_M^2 
\end{eqnarray}
with $\eta =Q^2/4M_d^2$. The deuteron form factor $F_d(Q^2)$ is defined as
$F_d(Q^2)\equiv\sqrt{A(Q^2)}$.
\par
As mentioned in Sec.~I, it is helpful to study the reduced form factor
since the effects of nucleon compositeness(represented by $F_N$)
have been removed from it. Equation~(\ref{redu}) means
that the deuteron form factor $F_d(Q^2)$ can be factorized into two parts.
and reduced form factor $f_d(Q^2)$
can be regarded as the form factor of a composite of two point-like
nucleons. This factorization was obtained by assuming
\begin{equation}\label{wffactor}
\Psi_d = \psi_d^{\rm body} \times {\psi_N} \times{\psi_N}
\end{equation}
in a simple covariant model \cite{bj}. $\psi_N$ is the nucleon wave 
function and $\psi_d^{\rm body}$ is the usual two-body 
wave function of the deuteron. 
The equation of motion for ${\Psi}_d (x_i,{\bf k}_{\perp i})$ 
in the light-cone frame(LCF) is given by
\begin{equation}\label{eqmo}
[M^2-\sum_{i=1}^6 \frac{{\bf k}_{\perp i}^2+m_i^2}{x_i}] 
       \Psi_d(x_i,{\bf k}_{\perp i} )= \\
  \int [dy][d^2{\bf j}_{\perp} ] V(x_i,{\bf k}_{\perp i} ;
      y_j,{{\bf j}_{\perp}}_j ) 
      {\Psi}_d(y_j,{{\bf j}_{\perp}}_j ) \hskip 0.1in .
\end{equation}
The factorized form of the deuteron form factor can be got by 
substituting Eq.~(\ref{eqmo}) into Drell-Yan formula \cite{dy}:
\begin{equation}\label{dyfm}
F_d(Q^2)=\sum_{a=1}^6 e_a \int [dx][d^2{\bf k}_{\perp i} ] 
     \Psi_d^{\ast}(x_i,{\bf k}_{\perp i}+(\delta_{ia}-x_i){ \bf q_{\perp}}) 
\Psi_d(x_i,{\bf k}_{\perp i} ) \hskip 0.1in ,
\end{equation}
where ${\bf q}_{\perp}$ is absorbed by the a-th quark,
$q=(0,q^-, {\bf q}_{\perp} )$ and $Q^2={\bf q}_{\perp}^2$. 
Noting that the gluon is a color octet in the SU(3) color group, 
the single-gluon exchange between two color-singlet nucleons
is forbidden. Thus the real kernel calculation requires the inclusion of 
other components rather than two-nucleons. Brodsky and Ji \cite{bj} suggested 
a simple covariant model to incorporate the quark structure of the nucleon.
The hard kernel at large $Q^2$ was assumed to be the perturbative amplitude 
for the six quarks to scatter 
from collinear to the initial two-nucleon configuration
to collinear to the final two-nucleon configuration, where each nucleon has
roughly equal momentum.
They argued that the dominant configuration for this
recombination is the quark-interchange plus one-gluon exchange between two
nucleons.
Thus, roughly speaking, we can divide the kernel into two parts. One 
represents the 
interchange of quarks and the gluon exchange between two nucleons, 
which transfer about half of the transverse momentum of the 
virtual photon from the struck
nucleon to the spectator nucleon. Another part is the inner evolution
of two nucleons. The first part leads to the reduced form factor of the deuteron
and the latter leads to the form factors of two nucleons together with the 
factorized wave function mentioned above.
\indent
\par
The body wave function in Eq.~(\ref{wffactor}) can be written as
\begin{equation} \label{wf}
\Psi_d^{\rm body}(y,{\bf l}_{\perp} )= A exp[- \frac{1}{ 2{\alpha}^2 }
\frac{{\bf l}_{\perp}^2 + m_N^2 }{4 y_1 y_2} ]
\end{equation}
by using Brodsky-Huang-Lepage prescription \cite{bhl} from a harmonic
oscillator wave function, $A^{\prime}exp[-\frac{1}{2}\alpha^2 r^2]$, in the 
rest frame.
As mentioned above, the reduced form factor can be obtained
\begin{equation}\label{fd}
f_d(Q^2)=D \int [dx][dy]\phi_d^{\dag}(x,Q) t_H(x,y,Q) \phi_d(y,Q)\hskip 0.1in ,
\end{equation}
where D is a kinematic factor. The body distribution amplitude $\phi_d(x,Q)$
is defined by
\begin{equation}\label{da}
\phi_d(x,Q)=\int [d{\bf k}_{\perp}] \Psi_d^{\rm body}(x_i,{\bf k}_{\perp i})
\end{equation}
with $x_{1} + x_{2} =1$ and ${{\bf k}_{\perp}}_1 +{{\bf k}_{\perp}}_2 =0$. 
The kernel $t_H(x,y,Q)$ is dominated by
the quark-interchange plus one-gluon-exchange diagrams. Since the binding
energy of the deuteron is very small, the lowest Fock state($NN$ configuration)
is dominant.

\section*{III. QCD INSPIRED MODEL}
\indent
\par
For the deuteron case, the matrix elements of the electromagnetic current 
$J^{\mu}$ can be written in terms of three form factors as
\begin{eqnarray}\label{gs}
G_{ {\lambda}^{\prime} \lambda }^{\mu}
&= & \langle P^{\prime} \lambda^{\prime} | J^{\mu} | P \lambda \rangle 
\nonumber\\
&= & -\{ G_1(Q^2) { {{\bf \epsilon}}^{\prime} }^{\ast} \cdot 
 {\bf \epsilon} [ P^{\mu}+{ P^{\prime} }^{\mu} ]+
 G_2(Q^2)[{{\bf \epsilon}}^{\mu} { {{\bf \epsilon}}^{\prime} }^{\ast} \cdot {\bf q} 
             -{{\bf \epsilon}}^{ \prime\ast\mu } {\bf \epsilon} \cdot {\bf q} ] \nonumber\\
&\mbox{} & -G_3(Q^2){\bf \epsilon} \cdot {\bf q} {{\bf \epsilon}}^{\prime\ast}\cdot {\bf q}
( P^{\mu}+ P^{\prime\mu} )/(2M^2) \} 
\end{eqnarray}
with $Q^2=-q^2$, $q=P^{\prime}-P$, and ${\bf \epsilon}\equiv{{\bf \epsilon}}_{\lambda}$, 
${\bf \epsilon}^{\prime}\equiv{\bf \epsilon}_{ \lambda^{\prime} }$ are the initial and final
polarization vectors, respectively. $| P \lambda \rangle$ is an eigenstate of momentum
$ P $ and helicity $ \lambda $.
The Lorentz invariant form factors $ G_{i} $
are related to the charge, magnetic and quadrapole form factors \cite{acg}:
\begin{eqnarray}
G_E & = & G_1 + {2\over 3}\eta G_Q \hskip 0.1in , \nonumber\\
G_M & = & G_2 \hskip 0.1in ,\nonumber\\
G_Q & = & G_1 - G_2 + (1+\eta ) G_3 \hskip 0.1in .
\end{eqnarray}
\par
Perturbative QCD predicts \cite{bl} that the helicity-zero to zero matrix 
element $G_{00}^+$ dominates helicity amplitude at large $Q^2$ for lepton
scattering on the deuteron. In the standard LCF, defined by $q^+ =0$, $q_y =0$
and $q_x =Q$, we have the following relations approximately:
\begin{eqnarray}
G_2 &=& 2 G_1=\frac{2}{2 P^+ (2 \eta +1)} G_{00}^{+} \hskip 0.1in ,\\
G_3 &=& 0 
\end{eqnarray}
and
\begin{equation} G_E :G_M :G_Q=(1-\frac{2}{3}\eta ):2:-1 
\end{equation}
while $Q\gg {\Lambda}_{\rm QCD}$. 
Also, Calson and Gross \cite{cg} have shown that
the LCF helicity-flip amplitudes $G_{+0}^+$ and $G_{+-}^+$ are suppressed by 
factors of ${\Lambda}_{\rm QCD}/Q$ and 
$( {\Lambda}_{QCD}/Q )^2$, respectively.
\par
Equation~(\ref{fd}) shows that the reduced form factor $f_d(Q^2)$ is determined
by the body distribution amplitude $\phi_d(x,Q)$ and the kernel which can 
be obtained from the quark-interchange plus one-gluon-exchange diagrams 
between two nucleons. To represent these diagrams, a model can 
be built by introducing a vector boson(color singlet) with an effective mass 
$M_b$ \cite{bh2}.
The picture of this model is described by Eq.~(\ref{fd}), which implies
that the formulation is similar to that of meson form factor and 
$t_H$ can be computed in the one-boson exchange approximation,
replacing the gluon by a massive vector
boson and coupling constant $g_s$ by effective coupling constant $g_{\rm eff}$.
We suggest that the effective mass $M_b$ can be 
determined by the empirical scaling law \cite{bh,arnold}:
\begin{equation}
(1+\frac{Q^2}{m_0^2} )f_d (Q^2) = constant
\end{equation}
with $m_0^2=M_b^2=0.28$ GeV$^2$ .
\par
The hard scattering amplitude can be obtained by calculating the diagrams 
shown in Fig.~2 :
\begin{equation}\label{th}
t_H(x,y,Q)=\frac{4M^2 g_{eff}^2}{xy Q^2 +{M_b}^2- { (x-y) }^2 M^2}\cdot
\frac{1}{xQ^2+(\frac{1}{4}- {(1-x)}^2 M^2) }\hskip 0.1in ,
\end{equation}
where M is the deuteron mass, and we have taken the nucleon mass 
to be half of M.
The kinematic factor D is
\begin{equation} 
D=\sqrt{   1+\frac{4}{3}\eta +\frac{4}{3}{\eta}^2    } \hskip 0.1in .
\end{equation}

\section*{IV. NUMERICAL RESULTS }
\indent
\par
Substituting Eq.~(\ref{th}) into Eq.~(\ref{fd}), 
numerical analysis can be done with several parameters,
$\alpha$, $M_b$ and $\alpha_{eff}=g_{eff}^2/4\pi$, in the expression. 
From the empirical scaling law, $M_b$ is around 0.5 GeV.
Instead of inputting the parameter $\alpha_{\rm eff}$,
we normalize the amplitude at $Q^2=2.5$ GeV$^2$ data point.
The variation of the reduced form factor $Q^2 f_d(Q^2)$ vs $Q^2$,
with different values of $M_b$, is displayed in Fig.~3. The effective 
coupling constant $\alpha_{eff}$ increases as $M_b$ become larger.
By varying the parameter $\alpha$ in the wave function 
we can get different behaviors of $Q^2 f_d(Q^2)$ in the
intermediate energy region, $Q^2\geq 1$ GeV$^2$.
The corresponding effective coupling constant, which decreases as 
$\alpha$ become larger, is shown in Fig.~4.

\par
It is shown from fitting the data that $M_b = 0.5$ GeV, 
$\alpha = 0.21$ GeV and $\alpha_{eff} = 0.15$.
The result, comparing with the calculations with the
Paris potential and the experimental data \cite{arnold,pla}, is shown in Fig.~1.
Our results reveal that our model with the above parameters can explain 
the deuteron form factor well for $Q^2\geq 1$ GeV$^2$.

\section*{V. SUMMARY AND DISCUSSION}
\indent
\par
The scaling law of the reduced from factor suggest \cite{bh} that the
dominance of $G_{00}^+$ begins at $Q^2 \sim 1$ GeV$^2$. Thus one
can calculate $G_{00}^+$ to predict the reduced form factor in the 
intermediate energy region.

However, it is a very complicated problem to directly calculate 
$G_{00}^+$,
since there are over 300,000 diagrams and the evolution of the 
deuteron wave function leads to the dominance of hidden-color 
state contributions in the very large $Q^2$ region due to the 
gluon exchange in the kernel. In fact, 
Farrar, Huleihel and Zhang \cite{fhz}
found that hidden-color degrees of freedom in the deuteron wave function
might be important in order to fit the data. In this paper we have tried
to build a model to calculate the reduced form factor in the intermediate 
energy region, instead of doing a full QCD analysis. The point of this model 
is that the reduced form factor $f_d(Q^2)$ can be
evaluated in a way similar to the meson form factor. It
is determined by the body wave function $\phi_d(x,Q^2)$ and
a kernel with a massive boson exchange. Our results show 
that our prediction can fit the data well for
$Q^2 > 1$ GeV$^2$. To fit the data we have chosen :
(i) the effective gluon mass $M_b\sim 0.5$ GeV, which is consistent with 
the empirical law, 
(ii) the parameter in the deuteron wave function $\alpha=0.21$ GeV, and
(iii) the normalization at $Q^2=2.5$ GeV$^2$ data point,
which corresponds to an effective coupling constant 
$\alpha_{eff} = 0.15  $. 

\par
In addition, we restrict
ourselves to calculate the reduced form factor only in the intermediate 
energy region. One can't expect that $G_{00}^+$ dominates the
helicity amplitude in the low $Q^2$ region, say, $Q^2 < 1$ GeV$^2$. 
On the other hand this picture can't apply to the form factor at
very large $Q^2$ since the hidden-color state contributions
may be important in that region, where the full evolution of the 
six-quark wave function is involved.

\par
This model could be improved by taking into account the contributions of 
$G_{+0}^+$ and $G_{+-}^+$ in the low $Q^2$ region and the hidden-color
contributions in the very large $Q^2$ region. We believe this model can be
generalized to other light nuclei.
 
\vspace{0.5in}
\begin{center} {\Large{\bf Acknowledgement} } \end{center}
\par
The authors would like to thank Professor T. Huang for his
valuable discussions.
\par
This work was supported by National Science Foundation of China (NSFC)
and Grant No. LWTZ-1298 of Academia Sinica.

\newpage
\noindent
{\Large\bf Figure Captions}

\vskip 0.3in
\noindent
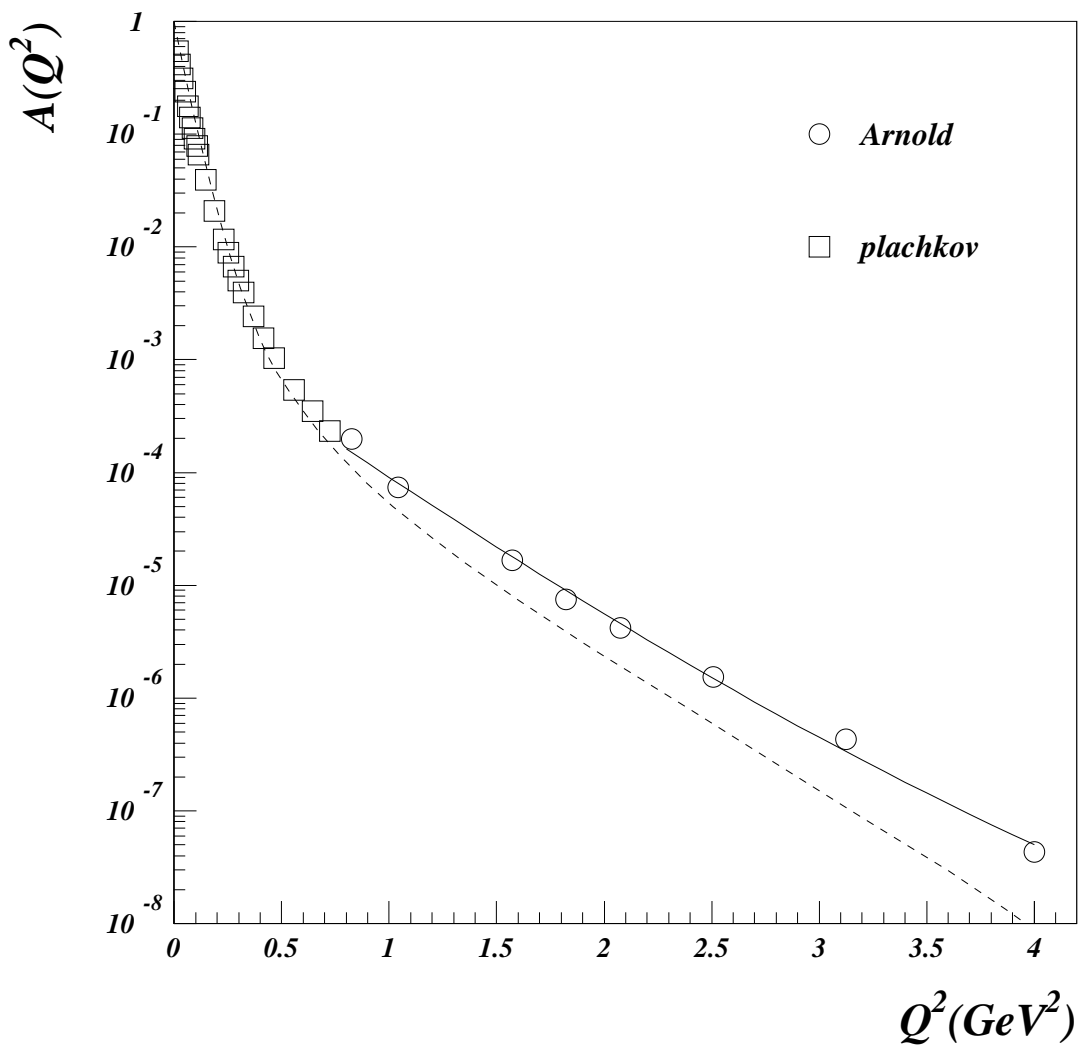
{\bf Fig.~1}
Structure function $A(Q^2)$ of the elastic $ed$-scattering from our
model(the solid line) with $M_b=0.5$ GeV, $\alpha = 0.21$ GeV,
and the effective coupling constant $\alpha_{eff}=0.15$.
The dashed line corresponds to the Paris potential calculation.

\vskip 0.2in
\noindent
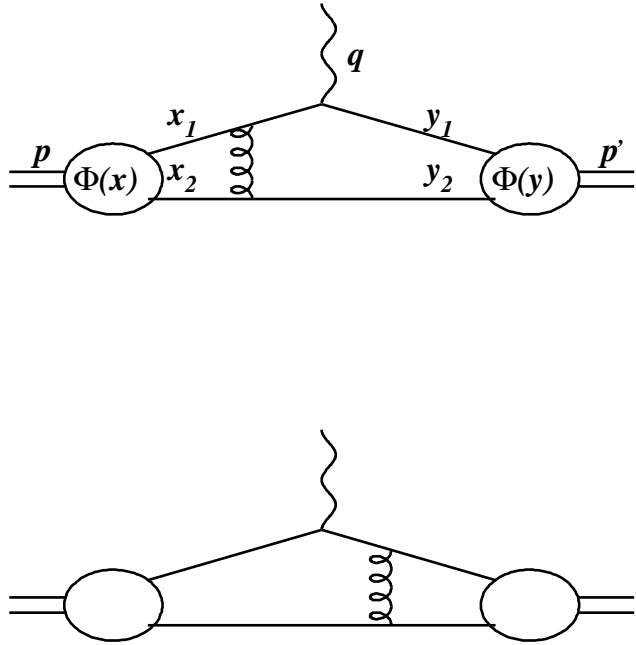
{\bf Fig.~2}
The hard scattering diagrams.

\vskip 0.2in
\noindent
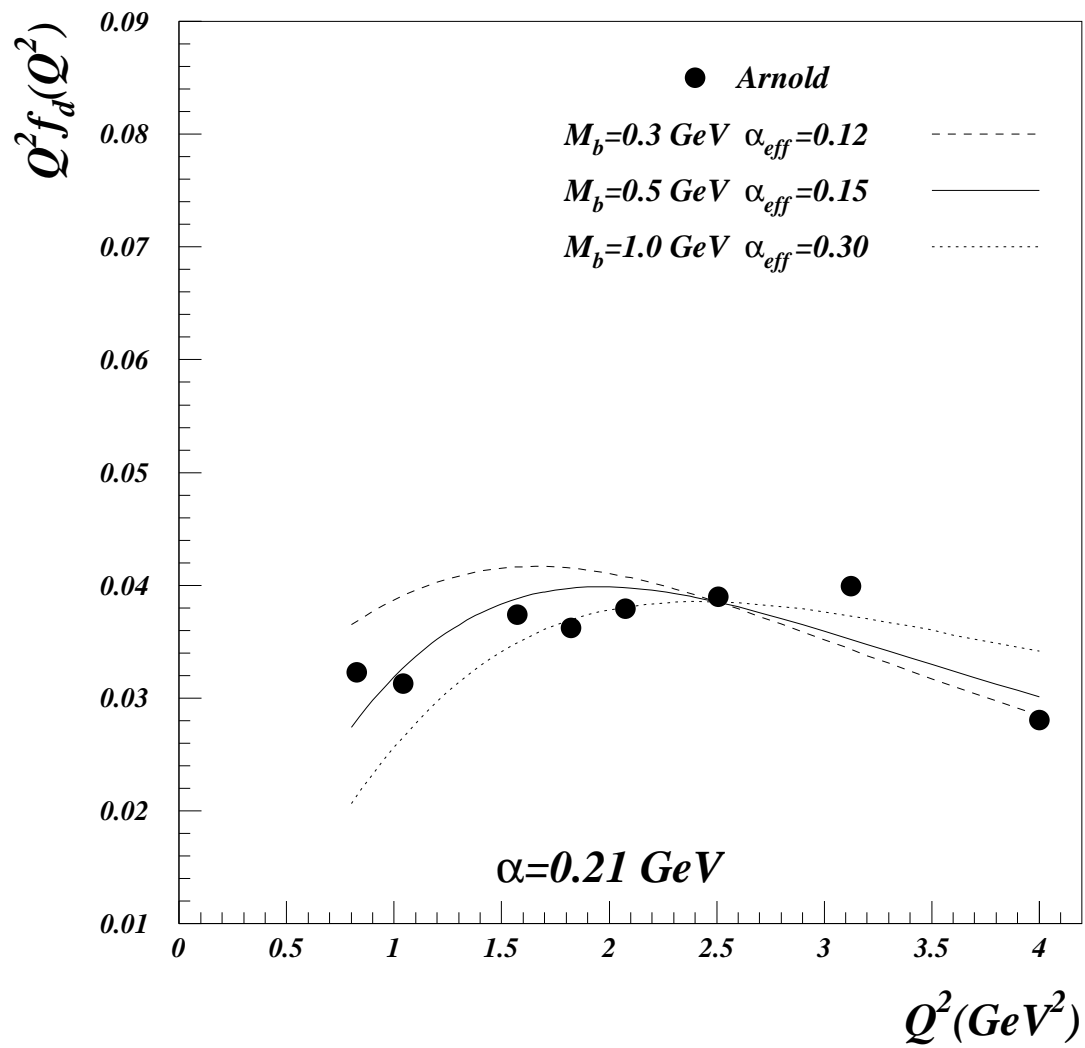
{\bf Fig.~3}
Comparison of the $Q^2 f_d(Q^2)$ data with our calculations by using
the different effective mass of the vector boson, $M_b$, while 
fixing $\alpha$ at $0.21$ GeV and normalized at the 
$Q^2 = 2.5$ GeV$^2$ data point.

\vskip 0.2in
\noindent
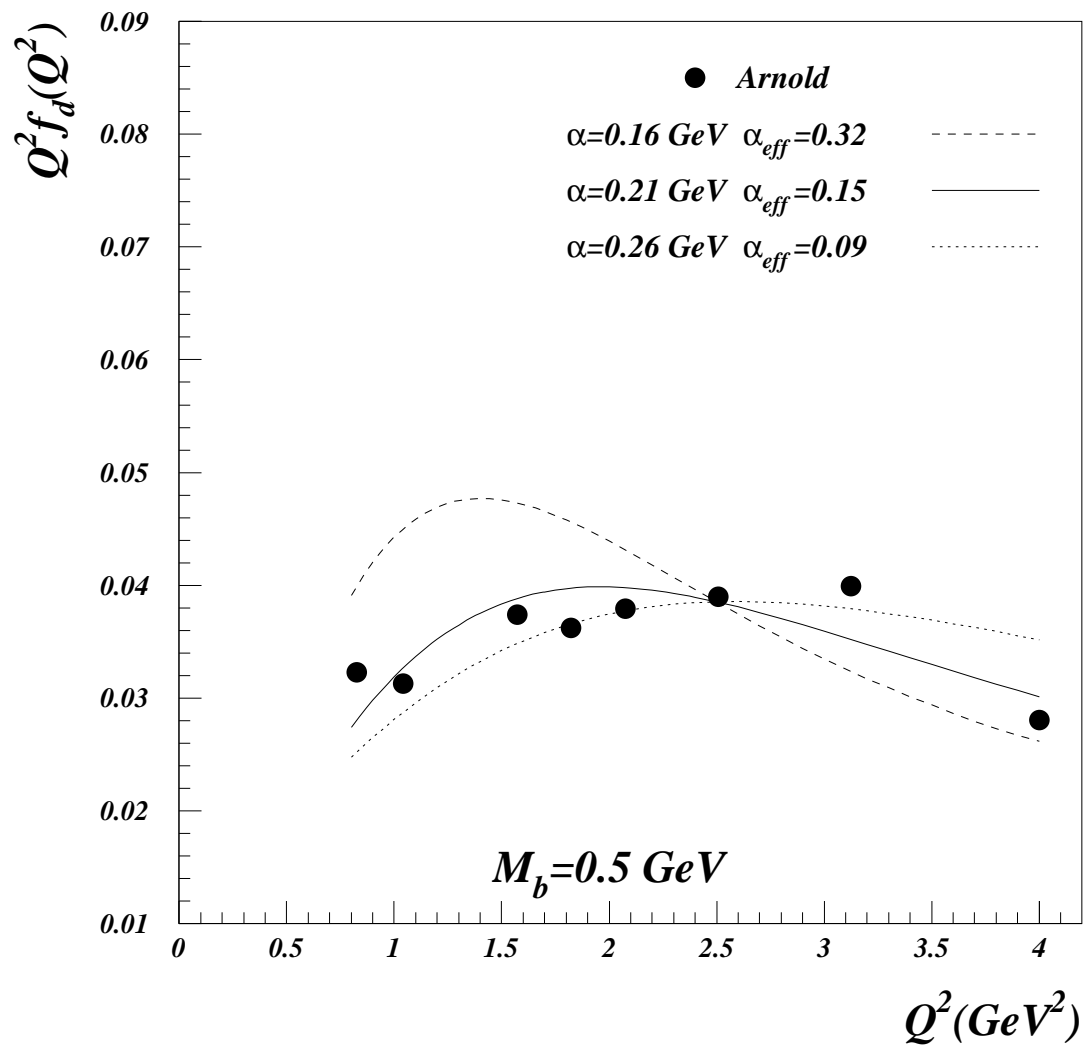
{\bf Fig.~4}
Comparison of the $Q^2 f_d(Q^2)$ data with our calculations by using
the different parameter in the wave function, $\alpha$, while fixing
$M_b$ at 0.5 GeV and normalized at the $Q^2 = 2.5$ GeV$^2$ data point.

\end{document}